\documentclass[runningheads,a4paper]{llncs}
\usepackage{cite}
\usepackage{amssymb}
\usepackage{graphicx}
\usepackage[misc]{ifsym} 
\usepackage{nonfloat}

\usepackage{floatrow}  
\newfloatcommand{capbtabbox}{table}[][\FBwidth]

\usepackage{url}
\usepackage[colorlinks,urlcolor=blue]{hyperref} 
\urldef{\mailsa}\path|wangmengsd@stu.xjtu.edu.cn, liukeen@xjtu.edu.cn|
\urldef{\mailsb}\path|simon2431@126.com, liuwenqiangcs@gmail.com|
\urldef{\mailsc}\path|whu@nju.edu.cn, sen.wang@griffith.edu.au|
\urldef{\mailsd}\path|sen.wang@griffith.edu.au|   
\urldef{\mailse}\path|xueli@itee.uq.edu.au| 
\newcommand{\keywords}[1]{\par\addvspace\baselineskip
\noindent\keywordname\enspace\ignorespaces#1}

\begin{document}

\mainmatter  

\title{PDD Graph: Bridging Electronic Medical Records and Biomedical Knowledge Graphs via Entity Linking}

\titlerunning{PDD Graph: Bridging Electronic Medical Records and Biomedical}

%
%
\author{Meng Wang\textsuperscript{1}%
\and Jiaheng Zhang\textsuperscript{1} \and Jun Liu \textsuperscript{1(\Letter)} \and Wei Hu\textsuperscript{2}\and\\  Sen Wang\textsuperscript{3}\and Xue Li\textsuperscript{4}\and Wenqiang Liu\textsuperscript{1}}
\authorrunning{M.Wang et al.\\}

\institute{1. MOEKLINNS lab, Xi'an Jiaotong University, Xi'an, China\\
	2. State Key Laboratory for Novel Software Technology, \\Nanjing University, Nanjing, China\\
3. Griffith Universtiy, Gold Coast Campus, Australia\\
4. The Universtiy of Queensland, Brisbane, Australia\\
	\mailsa\\
}


%
%

\toctitle{PDD Graph: Bridging Electronic Medical Records and Biomedical Knowledge Graphs via Entity Linking}
\tocauthor{M.Wang et al.}
\maketitle

\begin{abstract}
Electronic medical records contain multi-format electronic medical data that consist of an abundance of medical knowledge. Facing with patient’s symptoms, experienced caregivers make right medical decisions based on their professional knowledge that accurately grasps relationships between symptoms, diagnosis and corresponding treatments. In this paper, we aim to capture these relationships by constructing a large and high-quality heterogenous graph linking patients, diseases, and drugs (PDD) in EMRs. Specifically, we propose a novel framework to extract important medical entities from MIMIC-III (Medical Information Mart for Intensive Care III) and automatically link them with the existing biomedical knowledge graphs, including ICD-9 ontology and DrugBank. The PDD graph presented in this paper is accessible on the Web via the SPARQL endpoint, and provides a pathway for medical discovery and applications, such as effective treatment recommendations.
\newline 

\textbf{Resource type:} Dataset

\textbf{Permanent URL:} \url{http://kmap.xjtudlc.com/pdd}
\end{abstract}

\section{Introduction}
Big data vendors collect and store large number of electronic medical records (EMRs) in hospital, with the goal of instantly accessing to comprehensive medical patient histories for caregivers at a lower cost. Public availability of EMRs collections has attracted much attention for different research purposes, including clinical research \cite{wang2003cost}, mortality risk prediction \cite{ghassemi2014unfolding}, disease diagnosis \cite{wang2016diagnosis}, etc. An EMR database is normally a rich source of multi-format electronic data but remains limitations in scope and content. For example, MIMIC-III (Medical Information Mart for Intensive Care III) \cite{johnson2016mimic} collected bedside monitor trends, electronic medical notes, laboratory test results and waveforms from the ICUs (Intensive Care Units) of Beth Israel Deaconess Medical Center between 2001 and 2012. Abundant medical entities (symptoms, drugs and diseases) can be extracted from EMRs (clinical notes, prescriptions, and disease diagnoses). Most of the existing studies only focus on a specific entity, ignoring the relationship between entities. Given clinical data in MIMIC-III, discovering relationship between extracted entities (e.g. sepsis symptoms, pneumonia diagnosis, glucocorticoid drug and aspirin medicine) in wider scope can empower caregivers to make better decisions. Obviously, only focusing on EMR data is far from adequate to fully unveil entity relationships due to the limited scope of EMRs.

\begin{figure*}[t]
	\centering
	\includegraphics[width=0.75\textwidth]{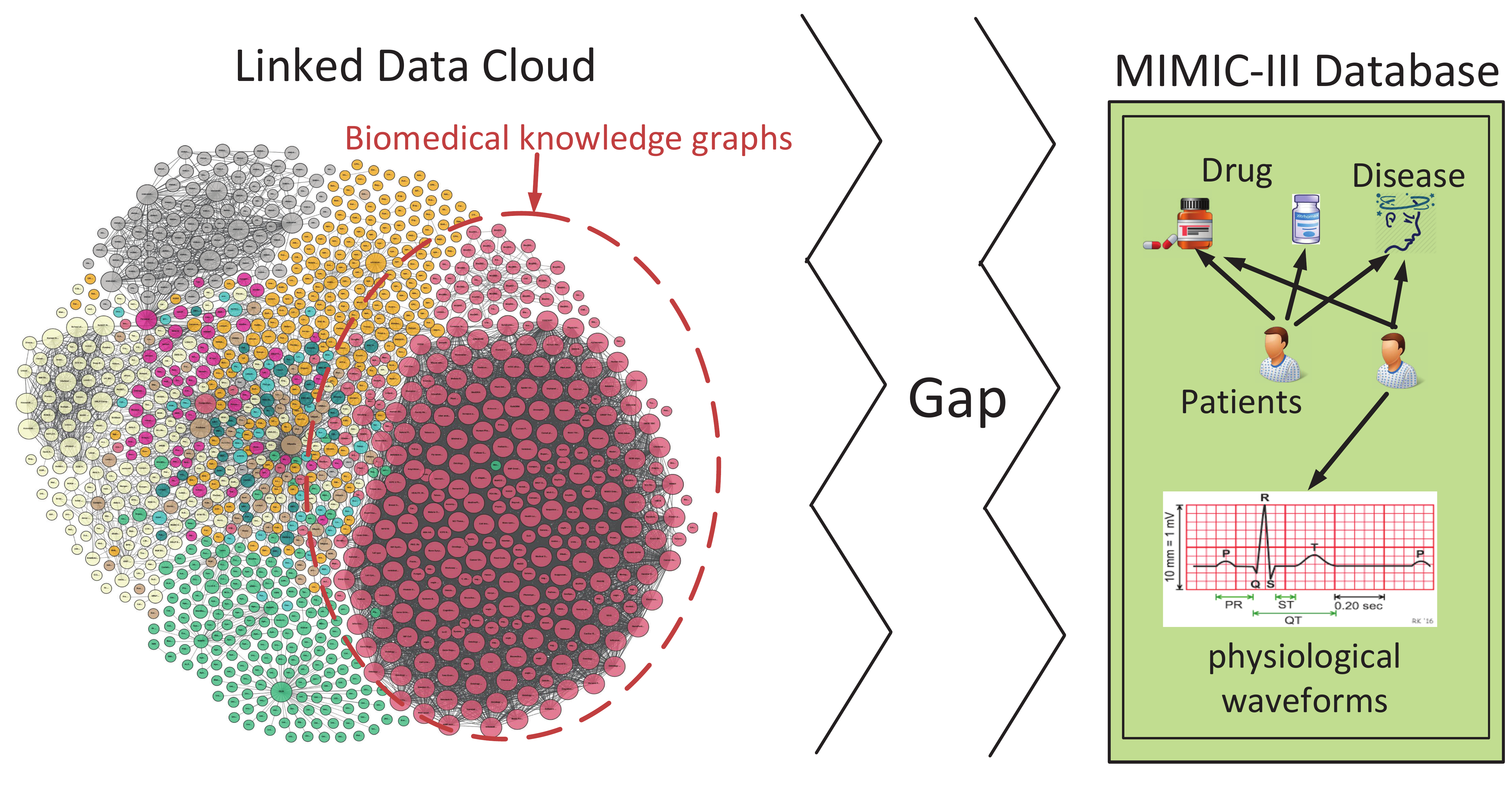}
		\vspace{-0.1in}
	\caption{Left part is the Linked Data Cloud$^1$, which contains interlinked biomedical knowledge graphs. Right part is the MIMIC-III database.}
	\label{intro}
\end{figure*} 

Meanwhile, many biomedical knowledge graphs (KGs) are published as Linked Data \cite{linkeddata} on the Web using the Resource Description Framework (RDF) \cite{rdf}, such as DrugBank \cite{drugbank} and ICD-9 ontology \cite{schriml2012disease}. Linked Data is about using the Web to set RDF links between entities in different KGs, thereby forming a large heterogeneous graph\footnote{Linking Open Data cloud diagram 2017. \url{http://lod-cloud.net/}}, where the nodes are entities (drugs, diseases, protein targets, side effects, pathways, etc.), and the edges (or links) represent various relations between entities such as drug-drug interactions. Unfortunately, such biomedical KGs only cover the basic medical facts, and contain little information about clinical outcomes. For instance, there is a relationship ``adverse interaction" between glucocorticoid and aspirin in DrugBank, but no further information about how the adverse interaction affect the treatment of the patient who took both of the drugs in the same period. Clinical data can practically offer an opportunity to provide the missing relationship between KGs and clinical outcomes. 

As mentioned above, biomedical KGs focus on the medical facts, whereas MIMIC-III only provides clinical data and physiological waveforms. There exists a gap between clinical data and biomedical KGs prohibiting further exploring medical entity relationship on ether side (see Fig.\ref{intro}). To solve this problem, we proposed a novel framework to construct a patient-drug-disease graph dataset (called PDD) in this paper. We summarize contributions of this paper as follows:
\begin{itemize}
	\item To our best knowledge, we are the first to bridge EMRs and biomedical KGs together. The result is a big and high-quality PDD graph dataset, which provides a salient opportunity to uncover associations of biomedical interest in wider scope. 
	\item We propose a novel framework to construct the PDD graph. The process starts by extracting medical entities from prescriptions, clinical notes and diagnoses respectively. RDF links are then set between the extracted medical entities and the corresponding entities in DrugBank and ICD-9 ontology. 
	\item We publish the PDD graph as an open resource\footnote{\url{http://kmap.xjtudlc.com/pdd}}, and provide a SPARQL query endpoint using Apache Jena Fuseki\footnote{\url{https://jena.apache.org/documentation/fuseki2/index.html}}. Researchers can retrieve data distributed over biomedical KGs and MIMIC-III, ranging from drug-drug interactions, to the outcomes of drugs in clinical trials.
\end{itemize}

It is necessary to mention that MIMIC-III contains clinical information of patients. Although the protected health information was de-identified, researchers who seek to use more clinical data should complete an on-line training course and then apply for the permission to download the complete MIMIC-III dataset\footnote{\url{https://mimic.physionet.org/}}.

The rest of this paper is organized as follows. Section 2 describes the proposed framework and details. The statistics and evaluation is reported in Section 3. Section 4 describes related work and finally, Section 5 concludes the paper and identifies topics for further work. 
	
	\vspace{-0.1in}
\section{PDD Construction}
	\vspace{-0.1in}
We first follow the RDF model \cite{rdf} and introduce the PDD definition.

\textbf{PDD Definition:} PDD is an RDF graph consisting of PDD facts, where a PDD fact is represented by an RDF triple to indicate that a patient takes a drug or a patient is diagnosed with a disease. For instance, 

 \centerline{$\langle$\textit{pdd}:274671, \textit{pdd}:diagnosed, sepsis$\rangle $\footnote{\textit{pdd} is the IRI prefix \url{http://kmap.xjtudlc.com/pdd_data/}}.}
\begin{figure}[ht]
	\centering
	\includegraphics[width=0.9\textwidth]{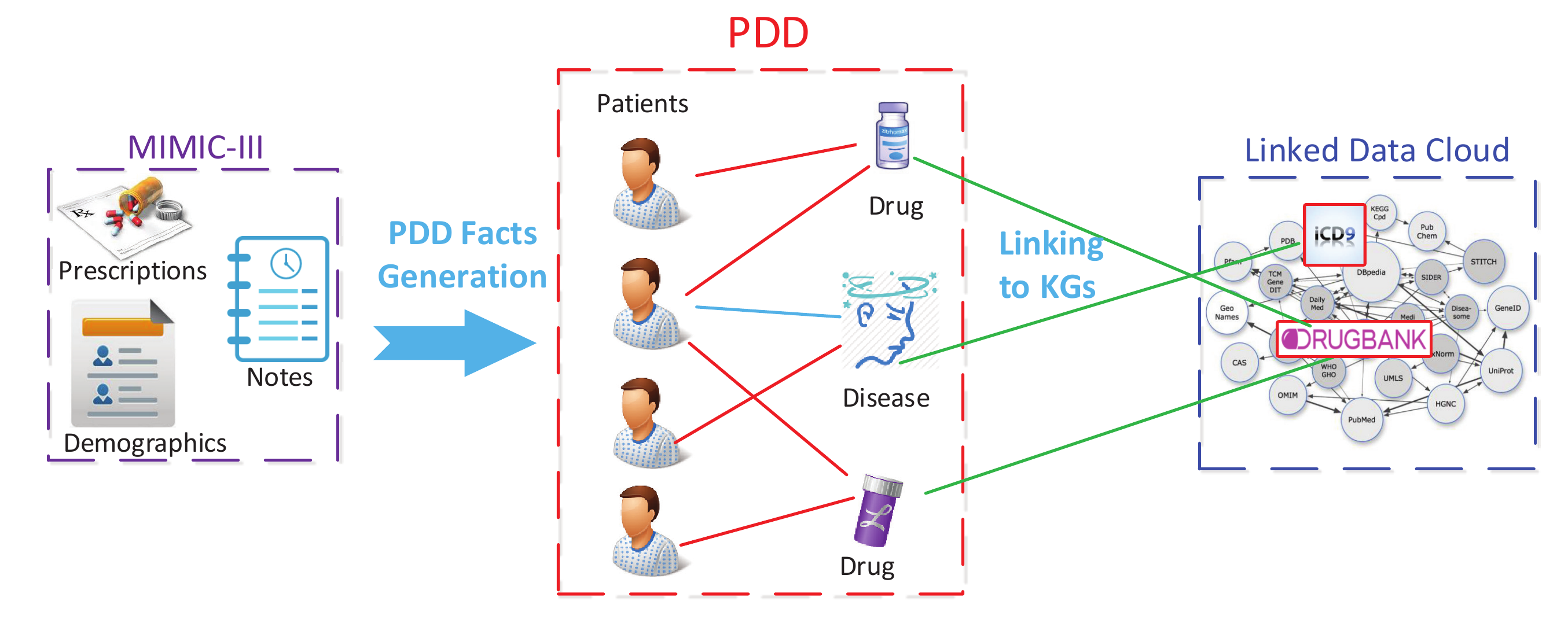}
	\vspace{-0.1in}
	\caption{Overview of PDD bridging MIMIC-III and biomedical knowledge graphs.}
	\label{overview}
\end{figure} 

Fig.\ref{overview} illustrates the general process of the PDD dataset generation, mainly includes two steps: PDD facts generation (described in Section 2.1), and linking PDD to biomedical KGs (described in Section 2.2).
\vspace{-0.1in}
\subsection{PDD Facts Generation}

According to the PDD definition, we need to extract three types of entities from MIMIC-III (patients, drugs, and diseases), and generate RDF triples of the prescription/diagnosis facts.

\textbf{Patients IRI Creation:} MIMIC-III contains 46,520 distinct patients, and each patient is attached with an unique ID. We add IRI prefix to each patient ID to form a patient entity in PDD.

\textbf{Prescription Triple Generation:} In MIMIC-III, the prescriptions table contains all the prescribed drugs for the treatments of patients. Each prescription record contains the patient's unique ID, the drug's name, the duration, and the dosage. We extracted all distinct drug names as the drug entities in PDD. Then we added a prescription triple in to PDD. An example is

 \centerline{$\langle$\textit{pdd}:18740, \textit{pdd}:prescribed, aspirin$\rangle $,}

where \textit{pdd}:18740 is a patient entity, and aspirin is the drug's name.

\begin{figure}[ht]
	\setlength{\abovecaptionskip}{0.cm}
	\setlength{\belowcaptionskip}{-0.5cm}
	\centering
	\includegraphics[width=0.6\textwidth]{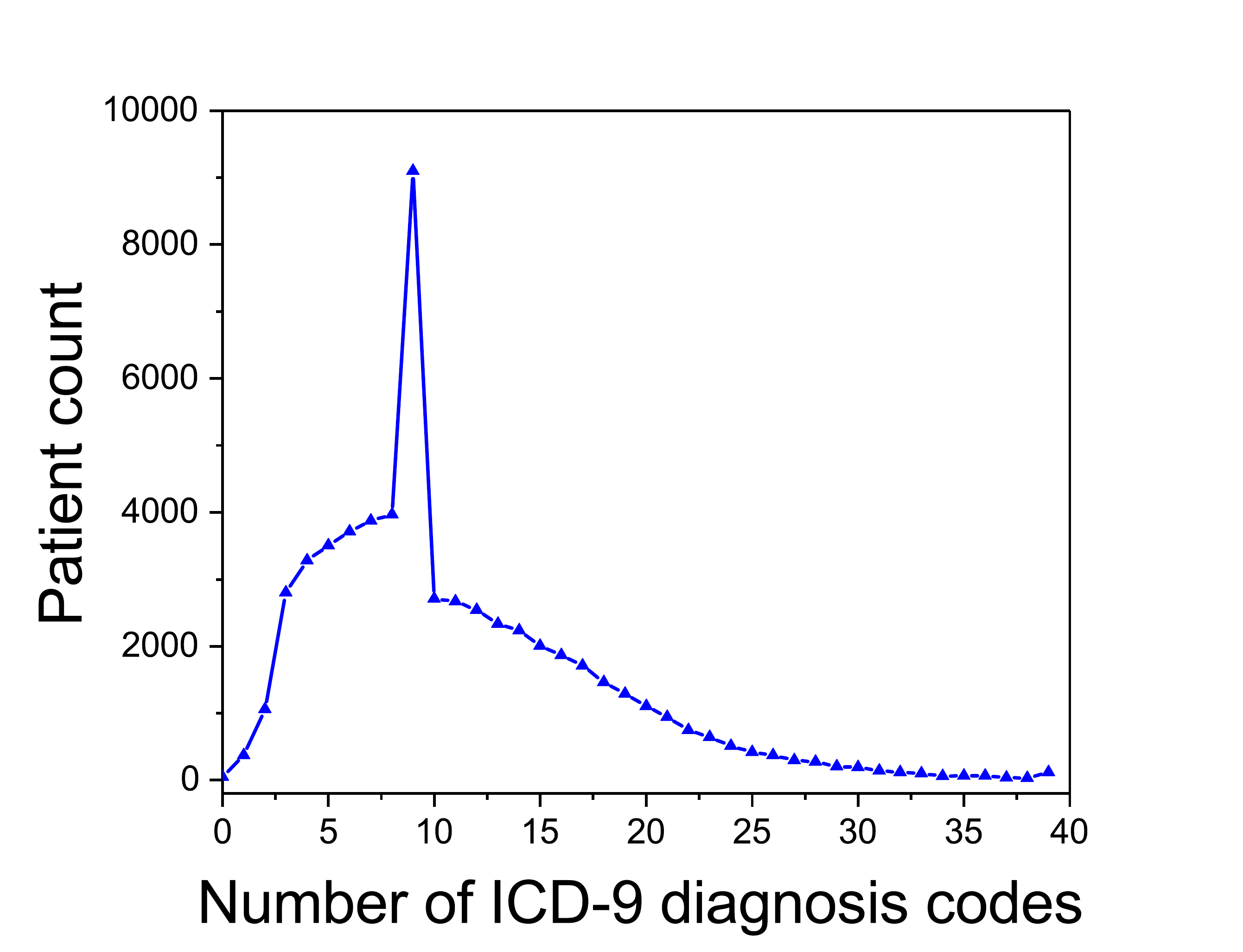}
	\caption{The distribution of assigned ICD-9 codes per patient.}
	\label{icd}
\end{figure}

\textbf{Diagnosis Triple Generation:} MIMIC-III provides a diagnosed table that contains ICD-9 diagnosis codes for patients. There is an average of 13.9 ICD-9 codes per patient, but with a highly skewed distribution, as shown in Fig. \ref{icd}. 
	\vspace{-0.4in}
	\vspace{-0.3in}
Beyond that, each patient has a set of clinical notes. These notes contain the diagnosis information. We use the named entity recognition (NER) tool C-TAKES \cite{savova2010mayo} to extract diseases from clinical notes. C-TAKES is the most commonly used NER tool in the clinical domain. Then we use the model \cite{wang2016diagnosis} (our previous work) to assign ICD-9 codes for extracted diseases. We extracted all ICD-9 diagnosis codes as the disease entities in PDD. Then we added a diagnosis triple into PDD. An example is

 \centerline{$\langle$\textit{pdd}:18740, \textit{pdd}:diagnosed, icd99592$\rangle $,}

where \textit{pdd}:18740 is a patient entity, and icd99592 is the ICD-9 code of sepsis.

\subsection{Linking PDD to Biomedical Knowledge Graphs}
After extracting entities, we need to tackle the task of finding \textit{sameAs} links \cite{ding2010sameas} between the entities in PDD and other biomedical KGs. For drugs, we focused on linking drugs of PDD to the DrugBank of Bio2RDF \cite{dumontier2014bio2rdf} version, as the project Bio2RDF provides a gateway to other biomedical KGs. Following the analogous reason, we interlinked diseases of PDD with the ICD-9 ontology in Bio2RDF. 

\textbf{Drug Entity Linking:} 
In MIMIC-III, drug names are various and often contain some insignificant words (10\%, 200mg, glass bottle, etc.), which challenges the drug entity linking if the label matching method is directly used. In order to overcome this problem, we proposed an entity name model (ENM) based on \cite{brown1993mathematics} to link MIMIC-III drugs to DrugBank. The ENM is a statistical translation model which can capture the variations of an drug's name.  

\vspace{-0.3in}
\begin{figure}[ht]
	\centering
	\includegraphics[width=0.5\textwidth]{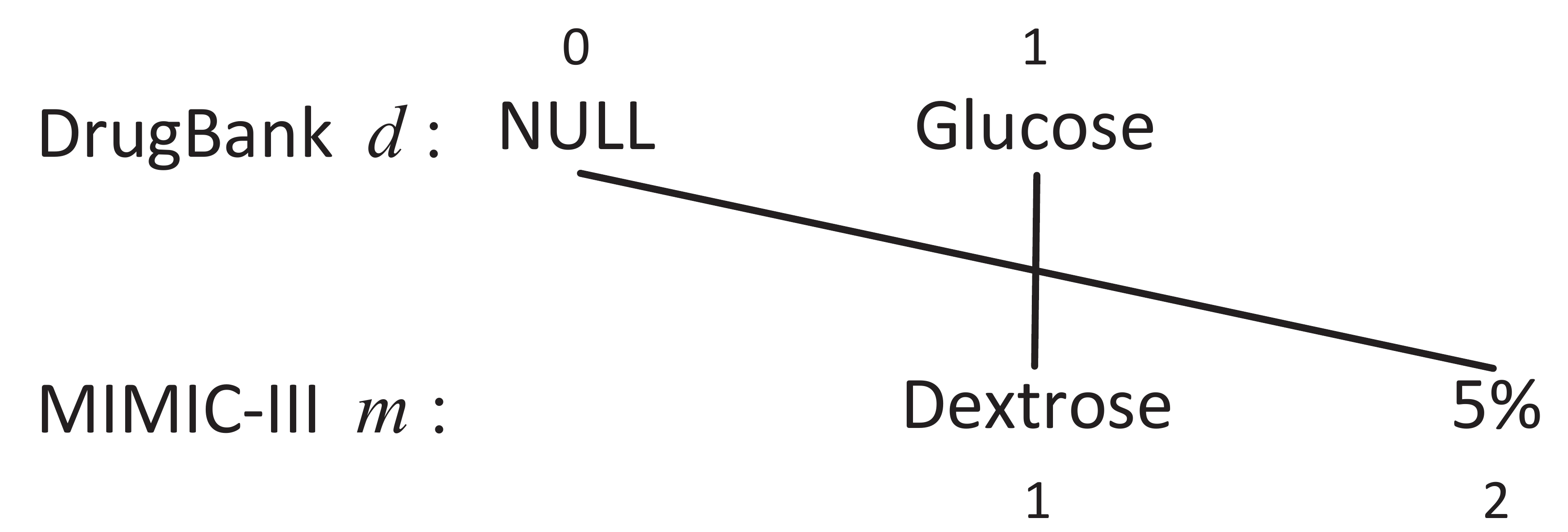}
	\vspace{-0.1in}
	\caption{The translation from \textit{Glucose} to \textit{Dextrose 5\%}.}
	\label{enm}
\end{figure} 
\vspace{-0.3in}

Given a drug’s name $m$ in MIMIC-III, the ENM model assumes that it is a translation of the drug’s name $d$ in DrugBank, and each word of the drug name could be translated through three ways:

1) Retained (translated into itself); 

2) Omitted (translated into the word NULL); 

3) Converted (translated into its alias).

Fig. \ref{enm} shows how the drug name \textit{Glucose} in DrugBank translated into \textit{Dextrose 5\%} in MIMIC-III. 

Based on the above three ways of translations, we define the probability of drug name $d$ being translated to $m$ as follows:
\vspace{-0.1in}
\begin{equation}\label{pm}
P(m|d)=\frac{\varepsilon}{(1_d+1)^{l_m}}\prod_{j=1}^{l_m}\sum_{i=0}^{l_d}t(m_i|d_j)
\end{equation}

where $\varepsilon$ is a normalization factor, $l_m$ is the length of $m$, $l_d$ is the length of $d$,  $m_i$ is the $i_{th}$ word of $m$, $d_j$ is the $j_{th}$ word of $d$, and $t(m_i|d_j )$ is the lexical translation probability which indicates the probability of a word  $d_j$ in DrugBank being written as $m_i$ in MIMIC-III. DrugBank contains a large amount of drug aliases information, which can be used as training sets to compute the translation probability $t(m_i|d_j )$. After training the ENM from sample data, a drug name in MIMIC-III will be more likely to be translated to itself or aliases in DrugBank, whereas the insignificant words tend to be translated to NULL. Hence, our ENM can reduce the effects of insignificant words for drugs entity linking.

In addition, we propose two constraint rules when selecting candidate drugs for $m$, and discard those at odds with the rules. 

\textit{Rule 1:} One of the drug indications in DrugBank must be in accordance with one of the diagnoses of the patients who took the corresponding drug in MIMIC-III at least .

\textit{Rule 2:} The dosage of a drug that patients took in MIMIC-III must be in accordance with one of the standard dosages listed in DrugBank.

Finally, we will choose the drug name $d$ in DrugBank for the given drug $m$ in MIMIC-III with maximal $P(m|d)$, and $d$ satisfies the two constraint rules.

\textbf{Disease IRI Resolution:} In our previous work \cite{wang2016diagnosis}, we have assigned ICD-9 disease codes for extracted disease entities. Since the ICD-9 code is the international standard classification of diseases, and each code is unique. We can directly link the ICD-9 codes of PDD to ICD-9 ontology by string matching.

\vspace{-0.2in}
\section{Statistics and Evaluation}
\vspace{-0.1in}
In this section, we report the statistics of PDD and make the evaluation on its accuracy . At present PDD includes 58,030 entities and 2.3 million RDF triples. 

\vspace{-0.2in}
\begin{table*}  \scriptsize\scriptsize
	\begin{floatrow}  
		\capbtabbox{  
			\begin{tabular}{p{1cm}|p{1.6cm}|p{1.8cm}}  
				\hline  
				 & \#Overall  & \#Drug/disease linked to KG\\  
				\hline  
				Patient &  46,520& \\\hline 
				Drug & 4,525 & 3,449 \\  \hline 
				Disease & 6,985& 6,983\\
				\hline   
			\end{tabular}  
		}{  \renewcommand{\captionlabelfont}{\scriptsize} 
			\caption{Statistics of Entities}  
			\label{tab:tb1}  
		}  
	
		\capbtabbox{  
		\begin{tabular}{p{2.2cm}|p{1.6cm}|p{1.8cm}}   
			\hline  
			&\#Overall & \#Drug/disease linked to KG\\  
			\hline  
			Demographics & 165,526& \\ \hline  
			Patients-Drugs &1,517,702&  1,259,702 \\  \hline 
			Patients-Diseases& 650,987& 650,939\\
			\hline  
		\end{tabular}  
		}{  \renewcommand{\captionlabelfont}{\scriptsize} 
			\caption{Statistics of RDF triples }  
			\label{tab:tb2}  
		} 
	\end{floatrow} 
  
\end{table*}  

\begin{figure}[!ht]
	\setlength{\abovecaptionskip}{0.cm}
	\setlength{\belowcaptionskip}{-0.5cm}
	\centering
	\includegraphics[width=0.85\textwidth]{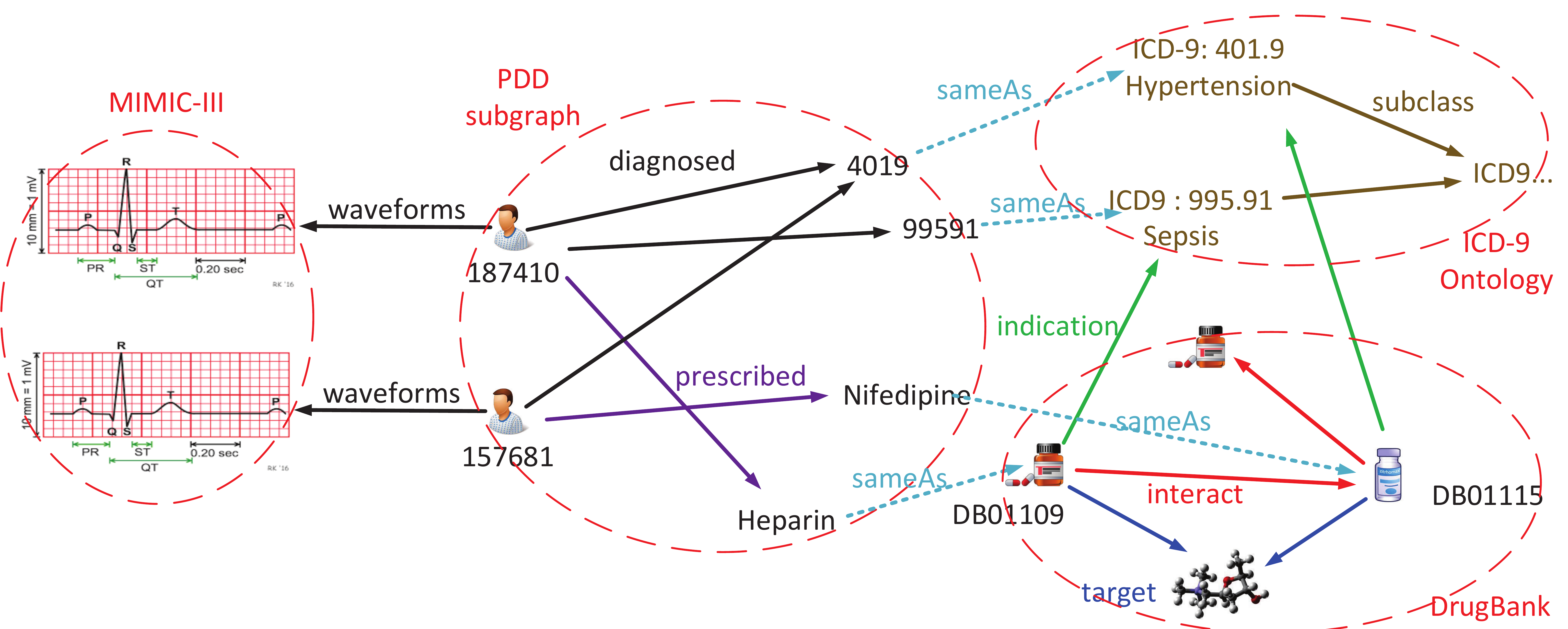}
	\vspace{-0.08in}
	\caption{An annotated subgraph of PDD.}
	\label{example}
\end{figure}

\vspace{-0.2in}
Table \ref{tab:tb1} shows the result of entities linked to the DrugBank and ICD-9 ontology. For drugs in PDD, 3,449 drugs are linked to 972 distinct drugs in DrugBank. For diseases in PDD, 6,983 diseases are connected to ICD-9 ontology. The only two failures of matching ICD-9 codes in MIMIC-III are '71970' and 'NULL', which are not included in ICD-9 ontology. Table \ref{tab:tb2} shows the result of RDF triples in PDD. In particular, 1,259,702 RDF triples contain drugs that have \textit{sameAs} links to DrugBank, and 650,939 RDF triples have ICD-9 diseases codes. It indicates 83.4\% drug-taken records in MIMIC-III can find corresponding entity in DrugBank, and 99.9\% diagnosed information can link to ICD-9 ontology. A subgraph of PDD is illustrated in Fig. \ref{example} to better understand the PDD graph.

\vspace{-0.3in}

\vspace{-0.25in}
To evaluate the ENM model, 500 samples are randomly selected, manually verified and adjusted. The ratio of positive samples to negative samples is 4:1, where positive means the entity can be linked to DrugBank. The precision is 94\% and the recall is 85\%. For linked entities in PDD we randomly chose 200 of them and manually evaluated the correctness of them, and the precision of entity links is 93\% which is in an accordance with the result of our examples. The overall accuracy of entity linking will be affected by the performance of the entity recognition tool. No entity recognition tools so far can achieve 100\% accuracy. The average accuracy of C-TAKES (we used in this paper) is 94\%. Therefore, the overall precision and recall may be lower.

In order to find out why those 1,076 drugs have not been linked to DrugBank yet, we extract 100 of them that hold the highest usage frequency. The observation shows that most of them are not just contained in DrugBank. For instance, DrugBank does not consider NS (normal saline) as a drug, but PDD contains several expressions of NS (NS, 1/2 NS, NS (Mini Bag Plus), NS (Glass Bottle), etc.). For drugs wrongly linked to DrugBank, the names of those drugs are too short, e.g. ‘N’ i.e nitrogen. These short names provide little information and affect the performance of ENM directly. Also, the training data from DrugBank does not include the usage frequency of each drug name. That might lead to some inconsistence with applications in MIMIC-III and cause linking errors. 

\vspace{-0.1in}
\section{Related Work}

\vspace{-0.1in}
In order to bring the advantages of Semantic Web to the life science community, a number of biomedical KGs have been constructed over the last years, such as Bio2RDF \cite{dumontier2014bio2rdf} and Chem2Bio2RDF \cite{chen2010chem2bio2rdf}. These datasets make the interconnection and exploration of different biomedical data sources possible. However, there is little patients clinical information within these biomedical KGs. STRIDE2RDF \cite{odgers2015mining} and MCLSS2RDF \cite{pathak2012applying} apply Linked Data Principles to represent patientʼs electronic health records, but the interlinks from clinical data to existing biomedical KGs are still very limited. Hence, none of the existing linked datasets are bridging the gap between clinical and biomedical data.

\vspace{-0.1in}
\section{Conclusion and Future Work}

\vspace{-0.1in}
This paper presents the process to construct a high-quality patient-drug-disease (PDD) graph linking entities in MIMIC-III to Linked Data Cloud, which satisfies the demand to provide information of clinical outcomes in biomedical KGs, when previous no relationship exists between the medical entities in MIMIC-III. With abundant clinical data of over forty thousand patients linked to open datasets, our work provides more convenient data access for further researches based on clinical outcomes, such as personalized medication and disease correlation analysis. The PDD dataset is currently accessible on the Web via the SPARQL endpoint. In future work, our plan is to improve the linking accuracy of ENM model by feeding more data into its training system.

\vspace{-0.1in}
\section*{Acknowledgment}

\vspace{-0.1in}
This work is sponsored by The Fundamental Theory and Applications of Big Data with Knowledge Engineering under the National Key Research and Development Program of China with grant number 2016YFB1000903; National Science Foundation of China under Grant Nos.61672419, 61370019, 61532004, 61672420, and 61532015; MOE Research Center for Online Education Funds under Grant No.2016YB165; Ministry of Education Innovation Research Team No.IRT17R86.
\vspace{-0.2in}

\vspace{-0.1in}
\bibliographystyle{splncs03}
\bibliography{reference}

\end{document}